\documentclass[preprint2]{aastex}

\input epsf.sty

\newcommand{\etal}{et~al.\ }
\newcommand{\eg}{e.g.\ }
\newcommand{\ie}{i.e.\ }
\newcommand{\Msun}{M_{\odot}}

\newcommand{\kms}{km~s$^{-1}$}

\newcommand{\Nifs}{$^{56}$Ni}
\newcommand{\Mej}{$M_{\rm ej}$}
\newcommand{\KE}{$E_{\rm kin}$}

\begin{document}

\title{The Type Ic Hypernova SN~2003dh/GRB~030329}

\author{
Paolo A.~Mazzali\altaffilmark{1,2,3},
Jinsong~Deng\altaffilmark{1,2},
Nozomu~Tominaga\altaffilmark{2}, 
Keiichi~Maeda\altaffilmark{2}, 
Ken'ichi~Nomoto\altaffilmark{1,2}, 
Thomas~Matheson\altaffilmark{4}, 
Koji~S.~Kawabata\altaffilmark{5},
Krzysztof Z. Stanek\altaffilmark{4},
Peter M. Garnavich\altaffilmark{6}
}

\altaffiltext{1}{Research Center for the Early Universe, School of Science, 
  University of Tokyo, Bunkyo-ku, Tokyo 113-0033, Japan}
\altaffiltext{2}{Department of Astronomy, School of Science, 
  University of Tokyo, Bunkyo-ku, Tokyo 113-0033, Japan;  
  deng@astron.s.u-tokyo.ac.jp, ntominaga@astron.s.u-tokyo.ac.jp, 
  maeda@astron.s.u-tokyo.ac.jp, nomoto@astron.s.u-tokyo.ac.jp}
\altaffiltext{3}{INAF-Osservatorio Astronomico, Via Tiepolo, 11,
  34131 Trieste, Italy; mazzali@ts.astro.it}
\altaffiltext{4}{Harvard-Smithsonian Center for Astrophysics, 60 Garden St., 
  Cambridge, MA 02138, USA; \\
  tmatheson@cfa.harvard.edu, kstanek@cfa.harvard.edu}
\altaffiltext{5}{Optical and Infrared Astronomy Division, NAOJ, Mitaka,
  Tokyo 181-8588, Japan; \\
  koji.kawabata@nao.ac.jp}
\altaffiltext{6}{Dept. of Physics, University of Notre Dame, 225
  Nieuwland Science Hall, Notre Dame, IN 46556, USA; \\
  pgarnavi@miranda.phys.nd.edu}

\begin{abstract}
The spectra of SN~2003dh, identified in the afterglow of GRB030329, are
modeled using radiation transport codes. It is shown that SN~2003dh had a high
explosion kinetic energy ($\sim 4 \times 10^{52}$\,erg in spherical symmetry),
making it one of the most powerful hypernovae observed so far, and supporting
the case for association between hypernovae and Gamma Ray Bursts. However, the
light curve derived from fitting the spectra suggests that SN~2003dh was not as
bright as SN~1998bw, ejecting only $\sim 0.35\Msun$ of \Nifs. The spectra of
SN~2003dh resemble those of SN~1998bw around maximum, but later they look more
like those of the less energetic hypernova SN~1997ef. The spectra and the
inferred light curve can be modeled adopting a density distribution similar to
that used for SN~1998bw at $ v > 25,000$\kms\ but more like that of SN~1997ef
at lower velocities. The mass of the ejecta is $\sim 8\Msun$, somewhat less
than in the other two hypernovae. The progenitor must have been a massive star
($M \sim 35-40\Msun$), as for other hypernovae. The need to combine different
one-dimensional explosion models strongly indicates that SN~2003dh was an
asymmetric explosion. 

\end{abstract}


\section{INTRODUCTION}

Evidence that at least some Gamma Ray Bursts (GRB's) are connected to
Supernovae (SNe) is mounting. After the serendipitous discovery of SN~1998bw in
coincidence with GRB980425 (Galama \etal 1998), several other cases of possible
SNe in GRB's have been reported (e.g., Bloom et al. 2002; Garnavich et al.
2003). All of these were however based only on the detection of `bumps' in the
GRB afterglows' light curves, which could be decomposed to look like the light
curve of SN~1998bw.

SN~1998bw was no ordinary SN. Its broad spectral features were explained as
indicating a very energetic Type Ic SN (arising from the collapse of the bare
CO core of a massive star).  Because of its high explosion kinetic energy
($\sim 5 \times 10^{52}$\,erg in spherical symmetry), and the consequently very
broad spectral lines, SN~1998bw was called a `hypernova' (Iwamoto \etal 1998).
Other hypernovae have been discovered and analyzed (\eg SNe 1997ef (Iwamoto
\etal 2000; Mazzali, Iwamoto, \& Nomoto 2000); SN~2002ap (Mazzali \etal 2002)),
but none was associated with a GRB. This may be related to the fact that none
of these SNe were either as bright or as powerful as SN~1998bw.  So far, the
only other positive detection of a SN spectrum in a GRB afterglow is the case
of GRB021211/SN~2002lt (della Valle \etal 2003). 

Given this intriguing but insufficient evidence, excitement mounted when a 
very nearby GRB was detected (GRB030329, z=0.1685; Greiner et al. 2003), as a
possible SN may be relatively easily observable in the light of the afterglow
(AG). 
Indeed, the detection of broad spectral features characteristic of 
a supernova was reported by Stanek \etal (2003) and later by 
Hjorth \etal (2003). The SN (SN~2003dh) appeared to be similar to SN~1998bw. 
However, spectra at $\sim 1$ month had changed somewhat, looking more like
those of the less energetic hypernova SN~1997ef (Kawabata \etal 2003). 

There are significant differences between the SN~2003dh light curves of
Matheson et al. (2003) and Hjorth et al. (2003). These may be due to the
different observational methods and to major difficulties with subtraction of
a) the underlying AG spectrum, which can change with time in an unknown way,
and b) the host galaxy background.  In particular, the Hjorth et al. (2003)
light curve rises much more rapidly, reaches a brighter peak, and then drops
much faster. The Matheson et~al. (2003) light curve, on the other hand, has a
slower rise, and resembles that of SN~1998bw.  Unfortunately, neither light
curve covers the likely time of peak, $\sim 12 $ -- 15 days after the GRB,
because this time coincided with full moon.  The absolute rest-frame $V$
magnitude at peak may have been between $-18.5$ and $-19.1$, depending on the
dataset used and the estimated extinction. Bloom et~al. (2003), however, find
that SN~2003dh may have been as bright as $M_V = -19.8 \pm 0.4$ at peak. They
used an extinction $A_V \approx 0.2$ toward SN 1998bw, explaining at least part
of the discrepancy. Also, they did not use spectral information in decomposing 
the light curve.

\section{Spectral Models}

Light curve models alone cannot uniquely constrain the properties of a SN, as
models yielding similar light curves may give rise to different synthetic
spectra (e.g. Iwamoto et al. 1998).  Fitting both light curves and spectra is a
much more effective approach. 

Unfortunately, in the case of SN~2003dh the exact shape of the light curve is
not yet certain. Therefore, we derived fiducial SN spectra by rescaling the
early power-law spectrum of the AG to the blue flux of the later spectra, where
a SN signature was evident, and subtracting it off. Attributing the entire blue
flux to the AG is justified by the fact that because of line-blanketing type
Ib/c SNe, like SNe Ia, always show a flux deficiency to the blue of $\sim
3600~\rm\AA$(e.g. Mazzali et al. 2000). At the third epoch the blue continuum
indicative of the AG is very weak, and no subtraction was applied. We fitted
three spectra of SN~2003dh using our Monte Carlo code (Mazzali \& Lucy 1993;
Lucy 1999; Mazzali 2000).  

The first spectrum was obtained at the MMT on 2003 April 10 (Matheson et al.
2003).  This is $\sim 12$ days after the GRB, \ie $\sim 10$ rest-frame days
into the life of the SN, assuming that the SN and the GRB coincided in time.
The spectrum is characterized by very broad absorption lines, and it is similar
to those of SN~1998bw at comparable epochs.  Using the same explosion model as
for SN~1998bw (model CO138E50; Table 1), a good match can be obtained for $\log
L = 42.83$ and $v(\rm ph) = 28000$\,\kms\ (Figure 1). The synthetic spectrum
has $V=-18.53$, $M(\rm Bol)=-18.37$. While the luminosity is lower than in
SN~1998bw, the photospheric velocity is comparable (for SN~1998bw we had $v(\rm
ph) = 31600$\,\kms on day 8 and 20700 on day 16).  This suggests that SN~2003dh
ejected a significant amount of high-velocity material in our direction, which
is understandable since we observed the GRB. SN~1998bw may have been even more
energetic, but is was probably viewed less on-axis since its associated GRB was
much weaker than GRB030329. Note that we cannot confirm the claim that
SN~2003dh has higher-velocity ejecta than SN~1998bw. The model used for
SN~1997ef (model CO100E18), on the other hand, has much too little mass at high
velocities, and it does not yield a broad-lined spectrum similar to the
observed one. 

The next spectrum we attempted to fit is the MMT 2003 April 24 one (Matheson et
al. 2003).  The spectrum (rest-frame epoch $\sim 23$ days) still has broad
lines, and it is almost as bright as the 10 April one, so that maximum probably
occurred about half way between the two epochs. If model CO138E50 is used, the
photosphere falls in the flat inner part of the density distribution (model
CO138E50 `turns over' at $v \sim 20000$\,\kms, which is above the position of
the photosphere at this epoch). The rather flat density distribution just above
the photosphere leads to very deep but narrow absorption lines, which are not
like the observed spectrum. On the other hand, model CO100E18 still has too
little mass at these velocities to give any significant spectral features for
such large velocities at this relatively advanced epoch (at a similar epoch
SN~1997ef had a low $v(\rm ph) = 10000$\,\kms). The spectrum is therefore in
the `transition' phase between SN~1998bw-like and SN~1997ef-like. Clearly, some
kind of new model is required. Since it is not clear how such a model should
behave, we deal first with the next epoch, which can yield indications about
the inner part of the ejecta, and then come back to the April 24 spectrum to
verify the results. 

Our third spectrum was obtained with Subaru on 2003 May 10 (Kawabata \etal
2003), a rest-frame epoch $\approx 36$ days, and it resembles that of SN~1997ef
at a comparable epoch. Indeed, model CO138E50 yields a narrow-lined spectrum,
while a much better synthetic spectrum is obtained using model CO100E18. This
model, in fact, turns over at $v \sim 6000$\,\kms, which is still below the
photosphere at this epoch. However, model CO100E18 is too massive at the low
velocities near the photosphere, resulting in strong backwarming and a high
temperature. We tested different possibilities, and found that a model where
the density of CO100E18 is divided by a factor of two throughout (CO100E18/2)
gives a good fit using $\log L = 42.43$ and $v(\rm ph) = 7500$\,\kms\ (Figure
2).  The synthetic spectrum has $V=-17.35$, $M(\rm Bol)=-17.37$. At a
comparable epoch the values for SN~1998bw were $\log L = 42.68$ and $v(\rm ph) =
7500$\,\kms, and those for SN~1997ef $\log L = 42.14$ and $v(\rm ph) =
4900$\,\kms. Therefore, it appears that while model CO138E50 is adequate for
the high velocity part of the ejecta, a less energetic model such as CO100E18/2
is better suited for the low-velocity inner part. This might be expected given
the shift in the properties of the spectrum. 

The question is how can those two models be merged into one. We have adopted
here an empirical approach. Model CO138E50 can be modified below $v =
28000$\,\kms, since this is below the photosphere of the April 10 spectrum,
while model CO100E18/2 can be modified at velocities high enough that the May
10 spectrum is not affected. We tried different boundaries for this, and
finally chose $v = 15000$\,\kms. Therefore, we used model CO100E18/2 for $v <
15000$\,\kms, model CO138E50 for $v > 25000$\,\kms, and merged the two models
linearly in between.  The density structures of the various models are shown in
Figure 4. With this new spherically symmetric `merged' model (COMDH) the results
for April 10 and May 10 are essentially unchanged.  COMDH has a smaller ejected
mass than both CO138E50 and CO100E18, \Mej $= 8 \Msun$, and  \KE $ = 3.8 \times
10^{52}$\,erg, similar to CO138E50 as it is dominated by the high-velocity
part. 
Since model COMDH is not computed from the hydrodynamics, 
but represents only a run of density vs. velocity, 
multi-dimensional calculations are necessary to see how well 
COMDH approximates a real multi-dimensional model.

The test for the merged model comes from fitting the April 24 spectrum, when
the photosphere falls in the joining region. Using essentially the same
parameters as discussed earlier ($\log L = 42.79$ and $v(\rm ph) =
18000$\,\kms), we could obtain a reasonably good fit to the observed spectrum
(Figure 3). The synthetic spectrum has $V=-18.17$, $M(\rm Bol)=-18.27$. 
Therefore, considering all uncertainties, our `merged' model is probably a 
fair one-dimensional representation of the density structure of SN~2003dh.

\section{A Two-Component Light Curve Model} 

In order to verify that the model we have constructed to reproduce the spectra
is indeed valid, it is necessary to compute a light curve with it. Here we can
follow the same approach as in the two-component models of Maeda \etal (2003),
as this is essentially the same problem.  Since we do not know the exact light
curve of SN~2003dh, we only tried to reproduce the three points derived from
the spectral fitting. These suggest a more rapid rise, a dimmer peak and a
faster decline than SN~1998bw, more similar to SN~2002ap. 

We computed synthetic bolometric light curves for models CO138E50 and COMDH,
using an LTE radiation transfer code and a gray $\gamma$-ray transfer code
(Iwamoto et al. 2000).  The optical opacity was found to be dominated by
electron scattering. 

The synthetic light curve of CO138E50 with homogeneous mixing of $^{56}$Ni out 
to 40,000 km s$^{-1}$ is in reasonable agreement with the three points 
of SN~2003dh inferred from the spectra (Figure 5).  
The mass of $^{56}$Ni ($0.32 \Msun$) is smaller than in the models for
SN~1998bw ($0.4-0.7 \Msun$: Nakamura et al. 2001; Iwamoto et al. 1998) and the
mixing more extensive. The model light curve of SN~2003dh rises faster and
peaks earlier than SN~1998bw.

The synthetic light curve computed with the `merged' model COMDH reproduces 
the three SN~2003dh points well for a $^{56}$Ni mass of 0.35$\Msun$.
Homogeneous mixing of $^{56}$Ni out to 40,000 km s$^{-1}$ was assumed, as for
CO138E50.  High velocity $^{56}$Ni is a feature that SN~2003dh shares with
other hypernovae. It could be attributed to a jet-like asphericity in the
explosion (Maeda \etal 2002; Maeda \& Nomoto 2003; Woosley \& Heger 2003).  

The main feature of the `merged' model COMDH is the presence of a high density
region at low velocities, as derived from the spectral analysis.  The effect of
this region should be to slow down the decline of the late-time light curve. In
SN~1998bw and other hypernovae this is indeed observed starting at $\sim 50$
days and becoming more prominent later (Sollerman \etal 2000; Nomoto \etal
2001; Nakamura \etal 2001; Maeda \etal 2003).  Figure 5 confirms that the
synthetic light curve of model COMDH becomes somewhat brighter than that of
CO138E50 at advanced phases.  Although the `merged' model is favored because it
gives reasonably good fits to both the light curve and the spectra, a
definitive extraction of the light curve and further time coverage will be
useful to distinguish between the models.

\section{Discussion} 

We have shown through spectral models that SN~2003dh was a hypernova. The high
ejecta velocities observed in the early spectra require for the outer layers an
explosion model similar to that used for SN~1998bw. However, at lower
velocities such a model becomes too flat, and it is necessary to replace it
with a lower energy one such as CO100, which has a steeper $\rho(v)$ at low
velocities. From a spherically symmetric calculation, we obtain for SN~2003dh
\KE $= 3.8 \times 10^{52}$\,erg, \Mej$ = 8 \Msun$, and $M$(\Nifs)$ = 0.35 \Msun$. 

Woosley \& Heger (2003) reached similar results. Their model parameters differ
from ours somewhat, but they did not make detailed comparisons with observed
data. 

Our results suggest that SN~2003dh was intermediate between SNe 1997ef
and 1998bw in \KE\ release and \Nifs\ production, and that it ejected a
comparable, possibly slightly smaller mass.  Accounting for the compact
remnant, the exploding star may have been a CO core of $\sim 10-11\Msun$,
which implies a main-sequence mass of $\sim 35-40\Msun$. 
SN~2003dh seems to follow the relations between progenitor mass, 
$M$(\Nifs) and \KE\ (Nomoto \etal 2003). 

All spherically symmetric models (including CO100 and CO138) are very flat
inside, and they all have an inner mass-cut defining a `hole' in the density
profile.  Evidence for SNe 1997ef, 2002ap and now 2003dh is that such a flat
density distribution does not lend itself to successful spectrum synthesis.
Actually, synthetic spectra computed for SN~1998bw beyond the epochs published
in Iwamoto \etal (1998) are also not perfect, so this may be a common feature. 
Evidence for a slow-moving, oxygen-dominated inner region in SN~2003dh, to be
obtained from the nebular spectra would nicely confirm this picture. 

The difficulties one-dimensional explosion models encounter in reproducing the
time evolution of hypernovae are almost certainly due to the inadequacy of such
models when applied to phenomena that are intrinsically aspherical. 
Approximate studies suggest that asymmetries in the explosions may affect the
light curves (H\"oflich, Wheeler, \& Wang 1999; Woosley \& Heger 2003). 
Therefore, the estimate of the explosion parameters may be subject to errors.
In particular, \KE\ may be overestimated by factors of a few by neglecting the
fact that the highest velocity material is only ejected in a narrow cone, as
discussed by Maeda \etal (2003), and recently by Woosley \& Heger (2003)
starting from the collapsar model. Both papers point at a similar scenario.
According to Woosley \& Heger (2003), if the high velocity ejecta are contained
in a cone with opening half-angle $\theta$, \KE\ is reduced by a factor
$(1-\cos \theta)$, and M(\Nifs) required to fit the peak luminosity is also
reduced by a factor $(1-\cos \theta)/\sin^2 \theta$ (in an extreme case).  This
is $\approx 0.5$  if $\theta$ is small.  The value of $M$(\Nifs) can be derived
from observations in the nebular phase, which is basically independent of
geometry.  At least for SN~1998bw, these confirmed that $M$(\Nifs) estimated in
spherical symmetry was correct within a factor of 2 or less (Nakamura \etal
2001; Maeda \etal 2002).  
Early-time spectra are less likely to be affected than the light curve, 
especially in the case of SN~2003dh, 
where the high velocity material which
dominates most spectral features is probably coming directly towards us.
Nebular observations of SN~2003dh are urged.

\acknowledgements This work was partly supported by the Japanese Ministry of
Education, Science, Culture, Sports, and Technology under Grants-in-Aid 
14047206, 14540223, 15204010.

\clearpage

\begin{deluxetable}{lrcc}
\tabletypesize{\scriptsize}
\tablecaption{Models}
\tablewidth{0pt}
\tablehead{
\colhead{Model} &
\colhead{$M_{\rm ej}$/$\Msun$} & 
\colhead{$E_{51}$} &
\colhead{$M$($^{56}$Ni)\tablenotemark{a}}\\
}
\startdata 
CO100E18\tablenotemark{b} &  9.6 & 18 & 0.32\\
CO138E50\tablenotemark{c} & 10.4 & 50 & 0.32\\
COMDH 			  &  8.0 & 38 & 0.35\\
\enddata 
\tablenotetext{a}{Values required to reproduce 
the peak brightness of SN~2003dh.}
\tablenotetext{b}{Mazzali et al. (2000).}
\tablenotetext{b}{Nakamura et al. (2001).}
\end{deluxetable}

\clearpage

\begin{figure*}
\centering
\epsscale{2.0}
\plotone{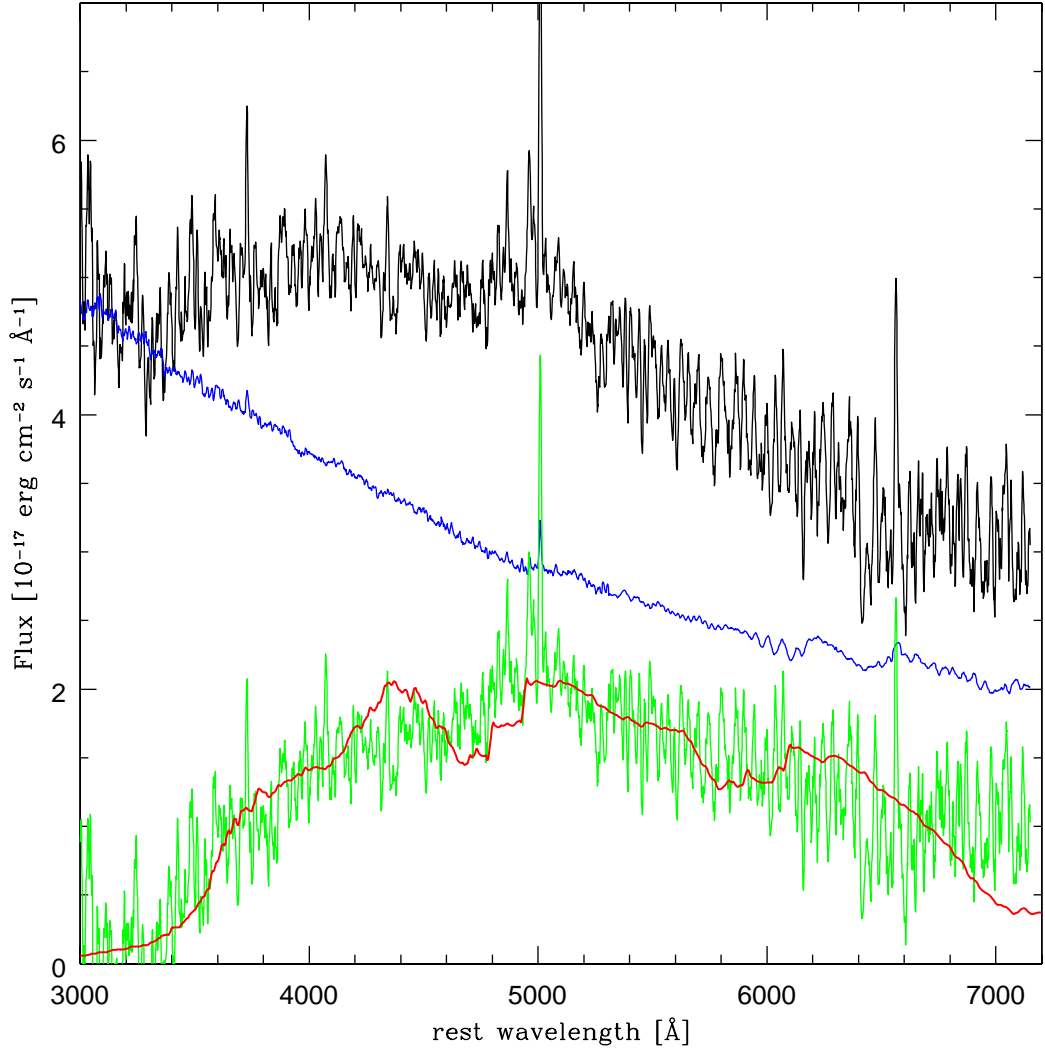}
\figcaption[sn2003dhapr10subcombco110div2co138.eps]
{The observed 2003 April 10 spectrum (top, black line); 
the subtracted afterglow spectrum (observed on 2003 April 1, middle, blue line); 
the `net' Supernova spectrum (bottom, green line), 
and our synthetic spectrum (bottom, red line).}
\end{figure*}
\clearpage

\begin{figure*}
\centering
\epsscale{2.0}
\plotone{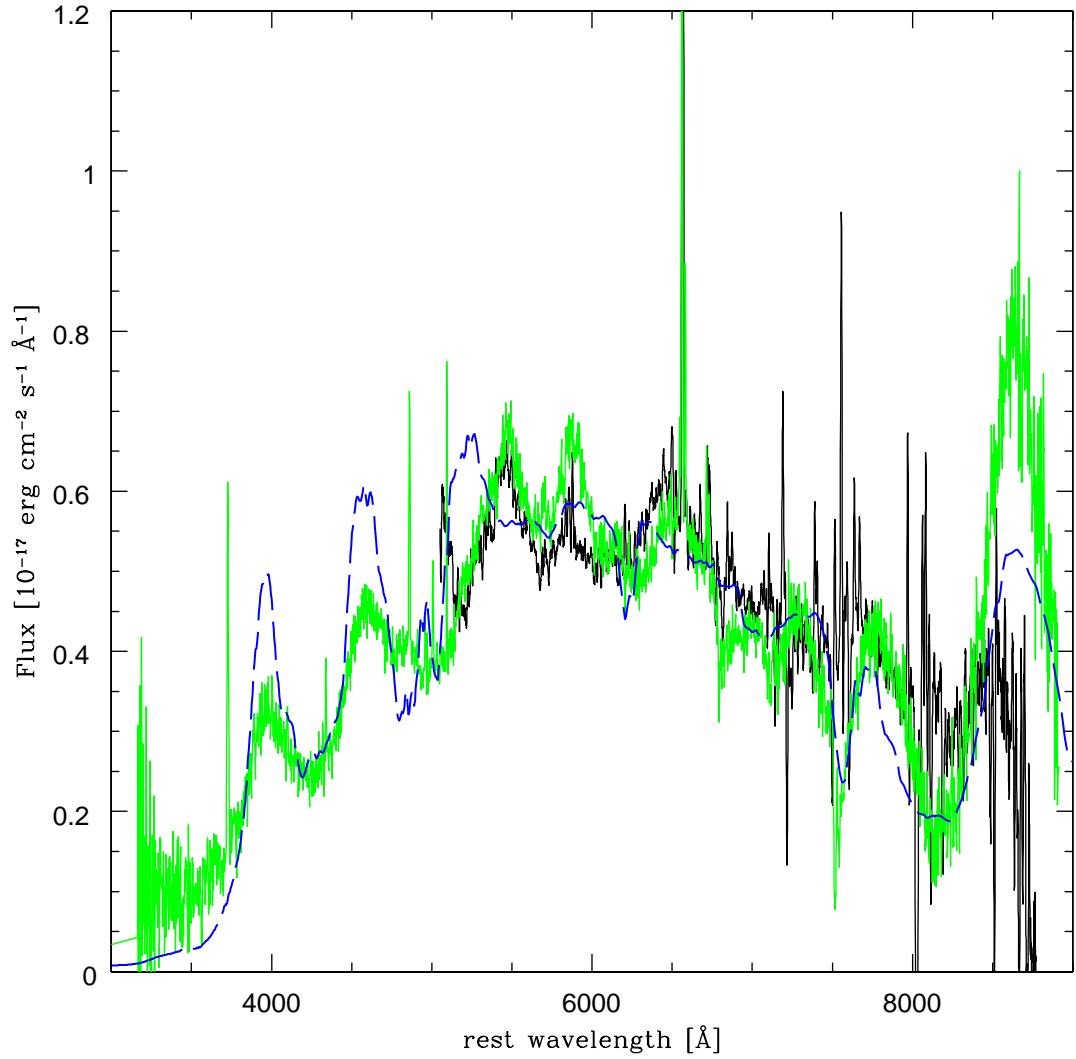}
\figcaption[sn2003dhmay10combco110div2co138.eps]
{The observed 2003 May 10 spectrum (black line); 
the spectrum of SN~1997ef obtained on 1998 January 1 (green line), 
and our synthetic spectrum (blue line).}
\end{figure*}
\clearpage

\begin{figure*}
\centering
\epsscale{2.0}
\plotone{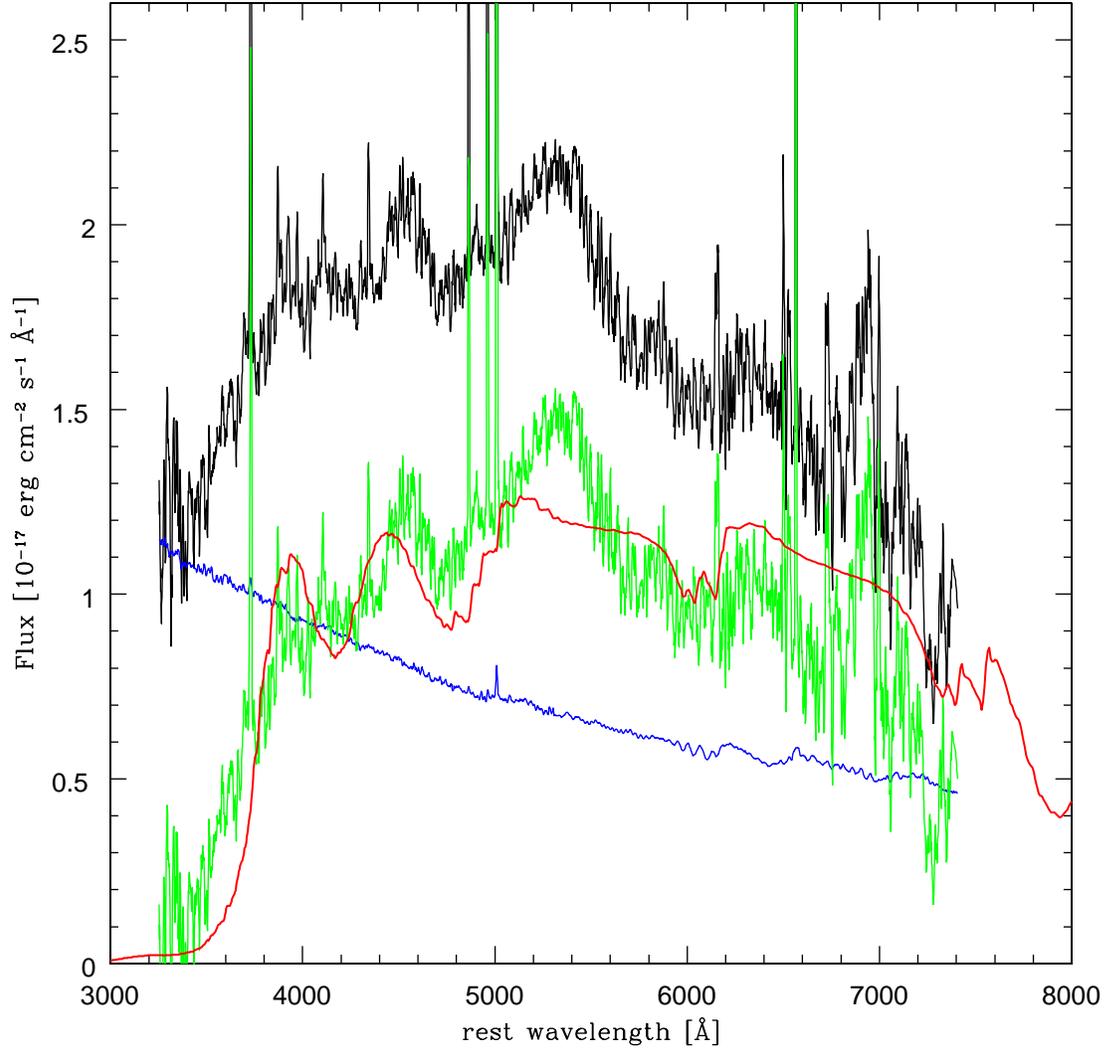}
\figcaption[sn2003dhapr24subcombco110div2co138.eps]
{The observed 2003 April 24 spectrum (top, black line); 
the subtracted afterglow spectrum (observed on 2003 April 1, bottom, blue line); 
the `net' Supernova spectrum (middle, green line), 
and our synthetic spectrum (middle, red line).}
\end{figure*}
\clearpage

\begin{figure*}
\centering
\epsscale{2.0}
\plotone{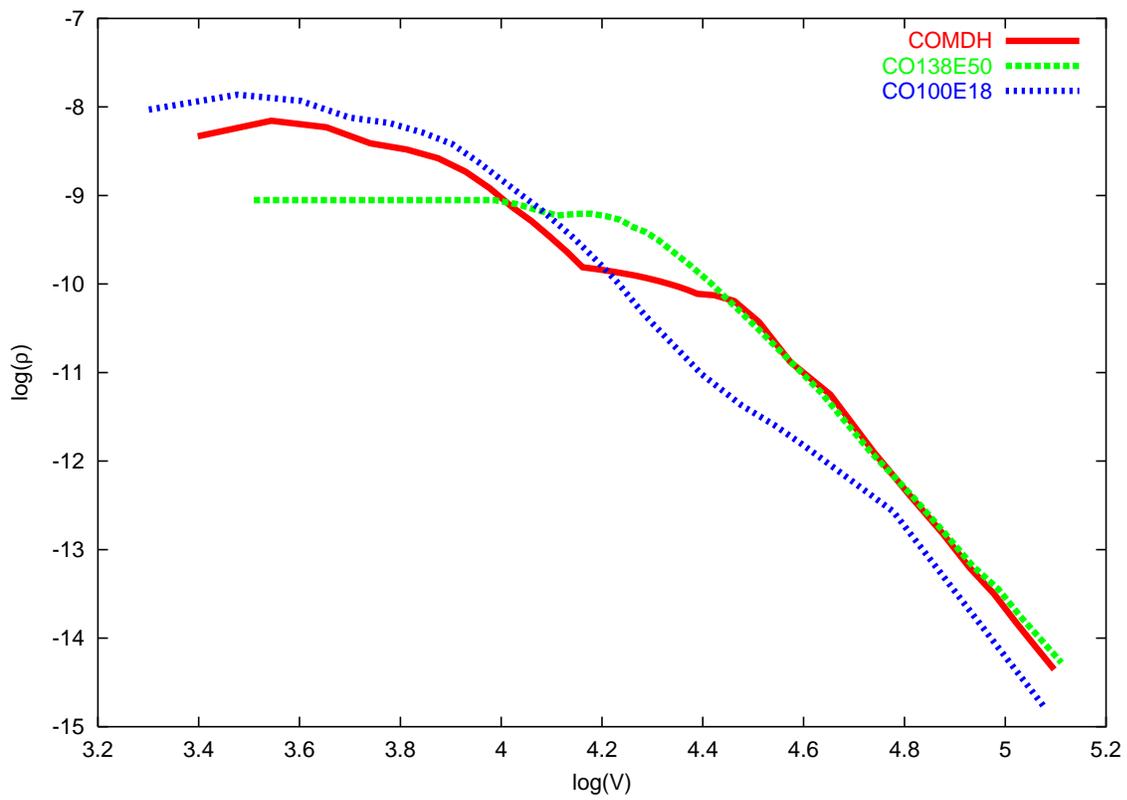}
\figcaption[sn2003dhdenprof.eps]
{Density profiles of the various models.}
\end{figure*}
\clearpage

\begin{figure*}
\centering
\epsscale{2.0}
\plotone{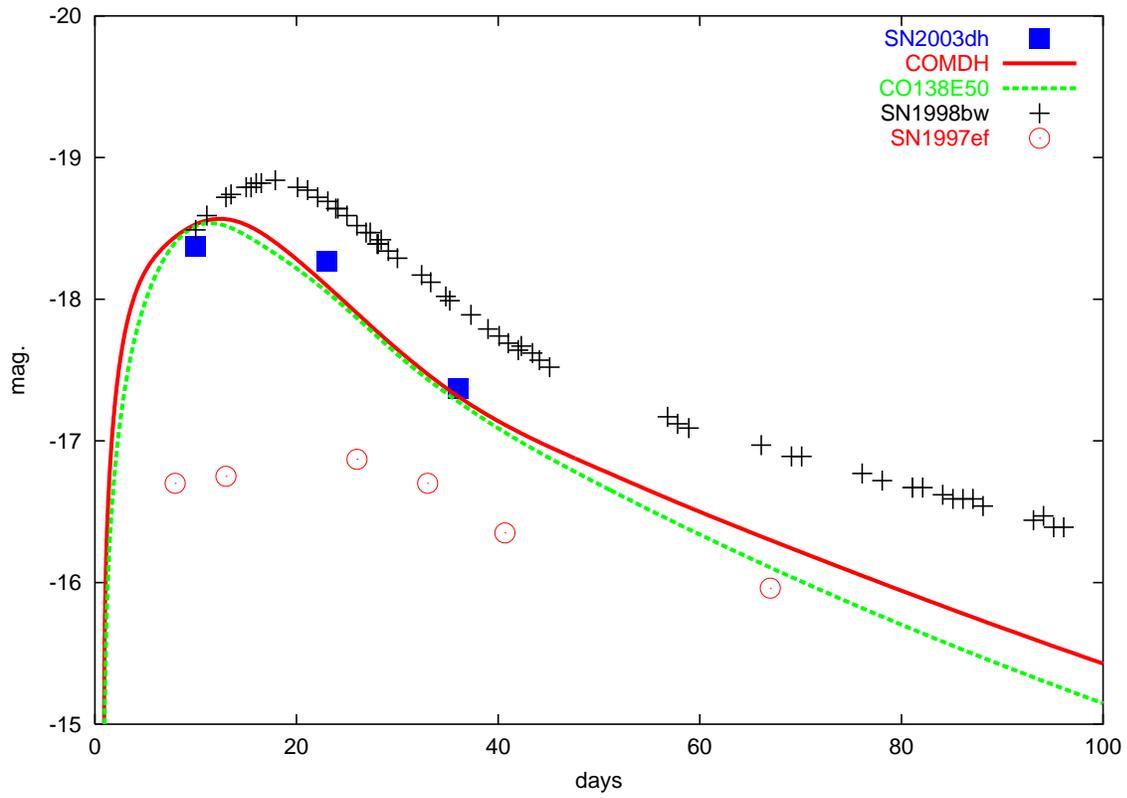}
\figcaption[sn2003dhlcurves.eps]{Observed and synthetic light curves. The bolometric light curve of
SN~1998bw is from Patat et al. (2001), that of SN~1997ef from Mazzali, Iwamoto, \& Nomoto(2000).}
\end{figure*}

\end{document}